\documentclass[12pt,english,aps,manuscript,aps,prl,superscriptaddress,showpacs,showkeys]{revtex4-1}
\usepackage[T1]{fontenc}
\usepackage[latin9]{inputenc}
\pdfoutput=1
\usepackage{geometry}
\usepackage{textcomp}
\usepackage{amstext}
\usepackage{graphicx}

\makeatletter

\providecommand{\tabularnewline}{\\}

\@ifundefined{textcolor}{}
{%
 \definecolor{BLACK}{gray}{0}
 \definecolor{WHITE}{gray}{1}
 \definecolor{RED}{rgb}{1,0,0}
 \definecolor{GREEN}{rgb}{0,1,0}
 \definecolor{BLUE}{rgb}{0,0,1}
 \definecolor{CYAN}{cmyk}{1,0,0,0}
 \definecolor{MAGENTA}{cmyk}{0,1,0,0}
 \definecolor{YELLOW}{cmyk}{0,0,1,0}
}

\usepackage{babel}

\makeatother

\usepackage{babel}
\begin{document}

\title{Grain boundary phases in bcc metals}

\author{T. Frolov}

\affiliation{Lawrence Livermore National Laboratory, Livermore, California 94550,
USA}

\author{W. Setyawan}

\affiliation{Pacific Northwest National Laboratory, P. O. Box 999, Richland, Washington
99352, USA}

\author{R. J. Kurtz}

\affiliation{Pacific Northwest National Laboratory, P. O. Box 999, Richland, Washington
99352, USA}

\author{J. Marian}

\affiliation{Department of Materials Science and Engineering, University of California
Los Angeles, Los Angeles, California 90095, USA}

\author{A. R. Oganov}

\altaffiliation[Current address: ]{Skolkovo Institute of Science and Technology, Skolkovo Innovation Center, 3 Nobel St., Moscow 143026, Russia}

\affiliation{Stony Brook University, Stony Brook, New York 11794, USA}

\author{R. E. Rudd}

\affiliation{Lawrence Livermore National Laboratory, Livermore, California 94550,
USA}

\author{Q. Zhu}

\affiliation{Department of Physics and Astronomy, High Pressure Science and Engineering
Center, University of Nevada, Las Vegas, Nevada 89154, USA}

\begin{abstract}
We report a computational discovery of novel grain boundary structures
and multiple grain boundary phases in elemental bcc tungsten. While
grain boundary structures created by the $\gamma$-surface method
as a union of two perfect half crystals have been studied extensively,
it is known that the method has limitations and does not always predict
the correct ground states. Here, we use a newly developed computational
tool, based on evolutionary algorithms, to perform a grand-canonical
search of a high-angle symmetric tilt boundary in tungsten, and we
find new ground states and multiple phases that cannot be described
using the conventional structural unit model. We use MD simulations
to demonstrate that the new structures can coexist at finite temperature
in a closed system, confirming these are examples of different GB
phases. The new ground state is confirmed by first-principles calculations. 
\end{abstract}
\maketitle
\bigskip{}

Grain boundaries play a key role in the mechanical behavior of body-centered
cubic (bcc) metals~\cite{Balluffi95}. Their structure affects a
diverse range of properties including fracture, recrystallization,
and creep~\cite{JACE:JACE03545,Luo23092011,tagkey2004xxi}. The abundance
of GBs in nanocrystalline materials makes the boundaries particularly
important in the class of advanced materials being developed for their
promise of radiation tolerance and the combination of ductility with
strength \cite{Lu349}. Beyond pure materials, alloy GB properties
are also significant, and the GB structure can have a marked effect
on segregation and morphological stability. The use of tungsten for
magnetic fusion energy applications provides a noteworthy example.
The ultimate success of fusion technology depends on materials that
can survive the harsh operating environment. Tungsten has been selected
as the divertor material in the International Thermonuclear Experimental
Reactor (ITER)~\cite{doi:10.1146/annurev-matsci-070813-113627} and
is a leading candidate for additional plasma-facing components in
the planned follow-on tokamak DEMO~\cite{0029-5515-47-11-014} because
of its high mechanical strength, high thermal conductivity, high melting
point and low yield for sputtering. While tungsten has a number of
favorable properties, it is also intrinsically brittle even at relatively
high temperatures especially after recrystallization \cite{Mutoh95}.
Recrystallization and intergranular fracture of tungsten impose significant
design constraints. Strategies that aim to improve the ductility include
alloying of the material with elements that improve the cohesion of
GBs. Small amounts of added elements may have a dramatic effect on
the fracture toughness of a material~\cite{Luo23092011,MillerMSEA02,Mutoh95,NiehSM84}.

One of the existing approaches to guide experimental alloy development
uses atomistic calculations based on density functional theory (DFT)
to screen the large space of elements in order to identify the best
candidates. These DFT simulations evaluate the propensity of the system
to undergo GB fracture by calculating the cleavage energy, which is
the difference $\gamma_{\text{GB}}-2\gamma_{\text{FS}}$ between the
boundary energy and the energy of the two free surfaces~\cite{RICE198923}.
Solutes segregating to different GB sites and free surfaces alter
the cleavage energy resulting in increased or decreased ductility,
provided the changes to the cleavage energy dominate over plasticity.
DFT calculations demonstrated that segregation to different boundary
sites may have opposing effects on the cleavage energy: segregation
to some sites improves GB cohesion, while segregation to other sites
promotes embrittlement \cite{SetSM12}. In order to accurately predict
the effect of segregation on embrittlement in these calculations it
is important to evaluate the cumulative effect of segregation to many
different sites. The energy of segregation to different GB sites strongly
depends on the local atomic environment, i.~e., the boundary structure~\cite{SetJPCM14,ScheiAM15,WuAM16}.
Atomistic simulations also demonstrated that the kinetics of crack
propagation is sensitive to grain boundary structure~\cite{PSSA:PSSA2211180112,Moller20141}.

Grain boundaries in bcc metals have been modeled using atomistic simulations
for several decades~\cite{MORITA19971053,PhysRevB.85.064108,doi:10.1080/13642818908211183,doi:10.1063/1.347741,PhysRevB.64.174101,Ratanaphan2015346}.
However, in a majority of the existing studies the 0 K temperature
structures were generated using the so-called $\gamma$-surface method.
In this approach two misoriented perfect half-crystals are joined
together while sampling different translations of the grains parallel
to the GB plane. The lowest energy GB configurations are assumed to
be the ground state in these calculations. During the minimization,
no atoms are added or removed from the GB core. In addition, the configurational
space of possible GB structures explored by the atoms during the energy
minimization is rather limited.

A number of computational studies in several different materials systems
demonstrated limitations of this approach and suggested that a more
thorough sampling that includes the optimization of the number of
atoms at the grain boundary is needed. For example, in ionic materials
low-energy grain boundary structures where found when a certain fraction
of ions was removed from the GB core prior to the energy minimization~\cite{doi:10.1080/01418618308243118,doi:10.1080/01418618608242811,DUFFY84a}.
Simulated quenching to the zero-temperature limit of the grand-canonical
ensemble demonstrated low-energy GB structures of a high-angle twist
grain boundary in face-centered cubic (fcc) Cu with different numbers
of atoms~\cite{Phillpot1992}. An investigation of Si twist boundaries
revealed the importance of sampling and optimization of the atomic
density and contrary to prior calculations demonstrated distinctly
ordered ground states at 0 K~\cite{Alfthan06}. Genetic algorithms
designed to explore a diverse population of possible structures were
applied to search for low-energy structures in symmetric tilt Si grain
boundaries~\cite{PhysRevB.80.174102} and multicomponent ceramic
grain boundaries~\cite{Chua:2010uq}.

Recent modeling in fcc metals showed that structure of relatively
simple GBs with high bicrystal symmetry can be surprisingly complex
and have multiple phases. Empirical potentials for Cu, Ag, Au and
Ni predicted new ground states and metastable phases of several {[}001{]}
symmetric tilt boundaries~\cite{Frolov2013,Frolov2013PRL,qzhutfrolov/online}.
These studies systematically explored GB energetics as a function
of number of atoms at the boundary. The new structures were found
to have different atomic densities and complex atomic ordering with
the periodic unit many times larger than that of the bulk crystals.
Improved simulation methodology demonstrated first-order reversible
transitions between grain boundary phases with different atomic densities
triggered by temperature, changes in chemical composition and concentration
of point defects~\cite{Frolov2013,qzhutfrolov/online,Frolov2016,PhysRevB.92.020103}.
In addition, the simulations showed that at certain temperatures and
chemical compositions GB phases can coexist in equilibrium with each
other~\cite{Frolov2013,Frolov2016}. Beyond symmetric tilt boundaries,
continuous vacancy loading into general grain boundaries in Cu revealed
lower energy states with different atomic density~\cite{Demkowicz2015}.

In bcc metals, recent computational studies demonstrated that changing
the number of atoms in the GB core increases~\cite{Novoselov2016276}
and in some boundaries decreases GB energy~\cite{Han2017,Han:2016aa}.
Little is know about phase behavior of grain boundaries in bcc metals
apart from the recently found dislocation pairing transition in low-angle
GBs composed of discrete dislocations~\cite{Olmsted2011}.

The $\gamma$-surface approach remains the most commonly used method
to construct GBs at 0 K, largely because no other robust computational
tool of GB structure prediction is available. On the other hand, much
progress has been made in developing of computational tools to predict
structures of crystals~\cite{Reilly:gp5080}. One such method is
USPEX~\cite{doi:10.1063/1.2210932}, which uses evolutionary algorithms
to predict the structure of materials based on the compositions alone.
USPEX has proved to be extremely powerful in different systems including
bulk crystals~\cite{doi:10.1063/1.2210932}, 2D crystals~\cite{Zhou-PRL-2014},
surfaces~\cite{Zhu-PRB-2013}, polymers~\cite{Zhu-JCP-2014} and
clusters~\cite{Lyakhov-CPC-2013}. In this work we use a recently
developed computational tool~\cite{qzhutfrolov/online} based on
the USPEX code~\cite{doi:10.1063/1.2210932,Lyakhov-CPC-2013,doi:10.1021/ar1001318}
to explore structures and phase behavior of GBs in a bcc metal.

In this work we reexamine the structure of the symmetric tilt $\Sigma27(552)[1\overline{1}0]$
grain boundary in tungsten which is a representative high-angle boundary
obtained by a 148$^{\circ}$ degree rotation of two grains around
a common $[1\overline{1}0]$ tilt axis. This choice of the model system
was motivated by several recent DFT studies that investigated $[1\overline{1}0]$
symmetric tilt boundaries to screen for alloying elements that would
improve ductility of tungsten~\cite{SetSM12,WuAM16,ScheiMSMSE16,Li14}.
In these studies the boundaries were constructed using the $\gamma$-surface
method. We performed simulations of the $\Sigma27(552)[1\overline{1}0]$
using two interatomic potentials, EAM1~\cite{Marinica2013} and EAM2~\cite{Zhou2001},
as well as DFT calculations~\cite{Kurtz2014,SetJPCM14}.

\begin{figure}
\begin{centering}
\includegraphics[width=0.8\paperwidth]{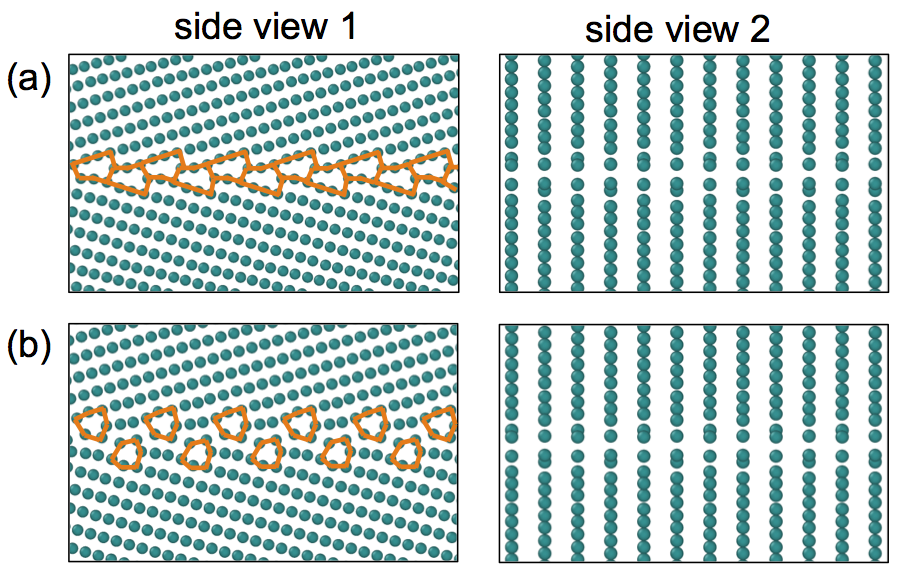} 
\par\end{centering}

\protect\protect\protect\caption{Conventional methodology: $\gamma$-surface constructed $\Sigma27(552)[1\overline{1}0]$
GB in W using (a) DFT \cite{SetJPCM14} and (b) the EAM1 \cite{Zhou2001}
and EAM2 \cite{Marinica2013} potentials. \label{fig:Kites_gamma_surface_approach}}
\end{figure}

First, the structure of the grain boundary was generated using the
$\gamma$-surface approach. The lowest energy configurations predicted
by DFT calculations in Ref.~\cite{SetJPCM14} and the current work
using the two interatomic potentials are illustrated in Fig.~\ref{fig:Kites_gamma_surface_approach}a
and b, respectively. The structural units of both configurations are
outlined with an orange curve to guide the eye. Notice that within
the $\gamma$-surface approach, DFT and the potentials predict somewhat
different structures. This result is consistent with previous studies
that identified multiple metastable and energy-degenerate states generated
by this methodology~\cite{Han2017}. The view of the boundary structure
along the tilt axis shown in the right-hand panel of Fig.~\ref{fig:Kites_gamma_surface_approach}
demonstrates that in both structures the atoms are confined to misoriented
(110) planes of the two crystals. By construction, these GB structures
can be mapped atom by atom onto a displacement symmetry conserving
(DSC) lattice~\cite{Balluffi95}, which is the coarsest lattice that
contains all of the atoms of both misoriented crystals on its lattice
sites.

Second, we constructed the boundary using an evolutionary search~\cite{qzhutfrolov/online}
as implemented in the USPEX code~\cite{doi:10.1063/1.2210932,Lyakhov-CPC-2013,doi:10.1021/ar1001318}.
In this approach a population of 50 to 100 different GB structures
evolves over up to 50 generations by operations of heredity and mutation
to predict low-energy configurations. The mutation operations include
the displacements of atoms, insertion and removal of atoms from the
GB core and sampling of larger-area GB reconstructions~\cite{qzhutfrolov/online}.

In our method, we split the bicrystal into three different regions,
the upper grain (UG), the lower grain (LG), and the grain boundary
(GB). UG and LG regions are taken to be 40~$\textup{\AA}$ thick.
The GB thickness is an input parameter predefined by the user. To
ensure accurate GB energy calculation converged with respect to the
system size normal to the grain boundary plane, we sandwich the GB
region between two 20~$\textup{\AA}$ thick buffer regions. The atoms
in the buffer zones are not affected by the evolutionary search, but
can move freely during the energy minimization. We create the first
generation of GB structures by randomly populating GB regions with
atoms, imposing layer group symmetries \cite{Lyakhov-CPC-2013} selected
at random for each bicrystal, and then joining the three regions together
applying random relative translations parallel to the grain boundary
plane. The enforced symmetry is used to avoid liquid-like structures
with close energies that are likely to produce similar children with
poor fitness. This initial symmetry can be broken or lowered by the
subsequent variation operations like heredity and mutation. The number
of atoms placed in each GB slab is estimated initially from the bulk
density of the perfect crystal and the thickness defined by the user.
This number is then randomly varied within the interval from 0 to
$N_{\text{plane}}^{\text{{bulk}}}$, where $N_{\text{plane}}^{\text{{bulk}}}$
is the number of atoms in one bulk atomic plane parallel to the GB.
This ensures that structures with different atomic densities are present
in the initial population. The atomic fraction {[}n{]} for each grain
boundary structure is calculated according to $[n]=(N\:\text{modulo}\: N_{\text{{plane}}}^{\text{{bulk}}})/N_{\text{plane}}^{\text{{bulk}}}$,
where $N$ is the total number of atoms in the GB region. We also
implemented constrained searches where {[}n{]} of all GB structures
in the population is within a certain interval. In the population
the different bicrystals have different GB dimensions generated as
random multiples of the smallest periodic GB unit \cite{Zhu-PRB-2015}.
The structures generated by the algorithm are relaxed externally by
the LAMMPS code \cite{Plimpton95} and the grain boundary energy is
determined and serves as a fitness parameter. During the optimization,
the atoms in the GB region need to be fully relaxed, while the atoms
in the bulk only move as rigid bodies.

Each successive generation is produced by operations of heredity and
mutations, by selecting the structures with the lowest 60\% of the
energies as parents. In the heredity operation two grain boundary
structures are randomly sliced and the parts from different parents
are combined to generate the offspring. In a mutation operation the
grain boundary atoms are displaced according to the stochastically
picked soft vibrational modes based a bond-hardness model \cite{Zhu-PRB-2015,Lyakhov-CPC-2013}.
Such mutations are advantageous to purely random displacements because
they mimic a structural transition due to phonon instability upon
large elastic strains and are more likely to lead to children structures
with low energy. To sample different atomic densities atoms in the
grain boundary region are inserted and deleted \cite{qzhutfrolov/online,Lyakhov-CPC-2013}.
The atoms are removed based on the value of the local order parameter
calculated for each atom. The order parameter is described in Eq.~(5)
of Ref.~\cite{LYAKHOV20101623}. To insert atoms into the GB slab,
we identify sites unoccupied by atoms by constructing a uniform grid
with a resolution of 1 \AA $^{3}$ and fill them at random. To ensure
relatively gradual changes in the GB structure, the random number
of the inserted and removed atoms also does not exceed 25\% of the
total number of atoms in the GB slab. A more detailed description
of the algorithm can be found in Ref.~\cite{qzhutfrolov/online}.

\begin{figure}
\begin{centering}
\includegraphics[width=0.8\paperwidth]{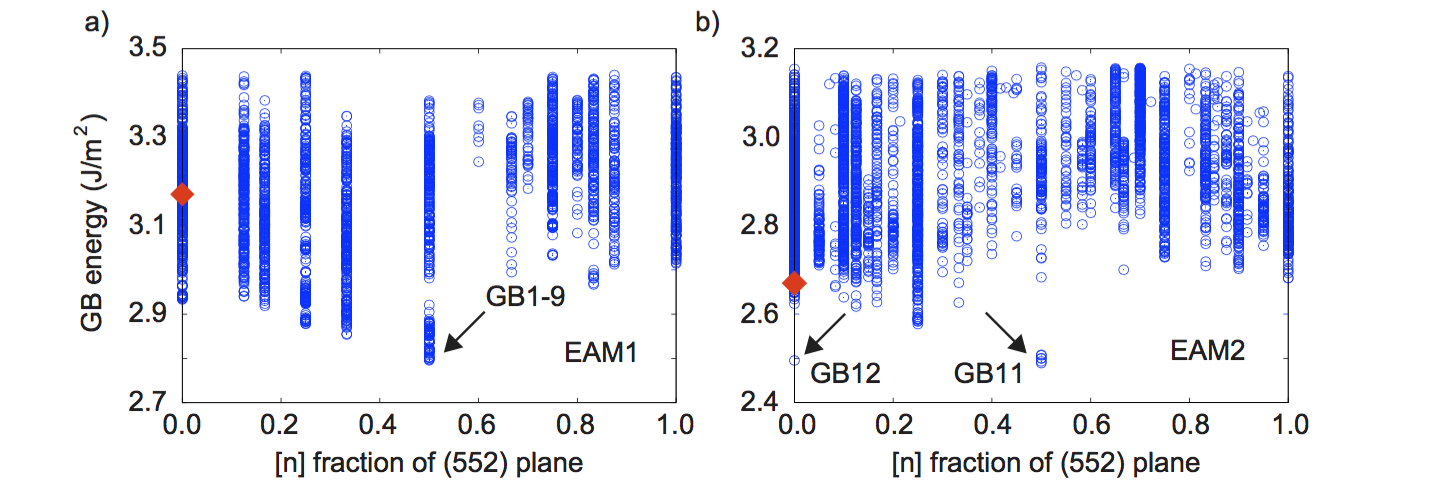} 
\par\end{centering}

\protect\protect\protect\caption{Results of the evolutionary search. The grain boundary energy for
different structures of the $\Sigma27(552)[1\overline{1}0]$ boundary
in W generated by the evolutionary search with (a) the EAM1 \cite{Marinica2013}
and (b) EAM2 \cite{Zhou2001} potentials. The energy is plotted as
a function of the number of atoms {[}n{]} expressed as a fraction
of atoms in the (552) plane. The red diamonds on the plots represent
the best configurations generated by the conventional $\gamma$-surface
approach. The arrows point to new ground states with different atomic
densities predicted by the evolutionary search. \label{fig:USPEX-search_GB_energy}}
\end{figure}

\begin{figure}
\begin{centering}
\includegraphics[width=0.7\paperwidth]{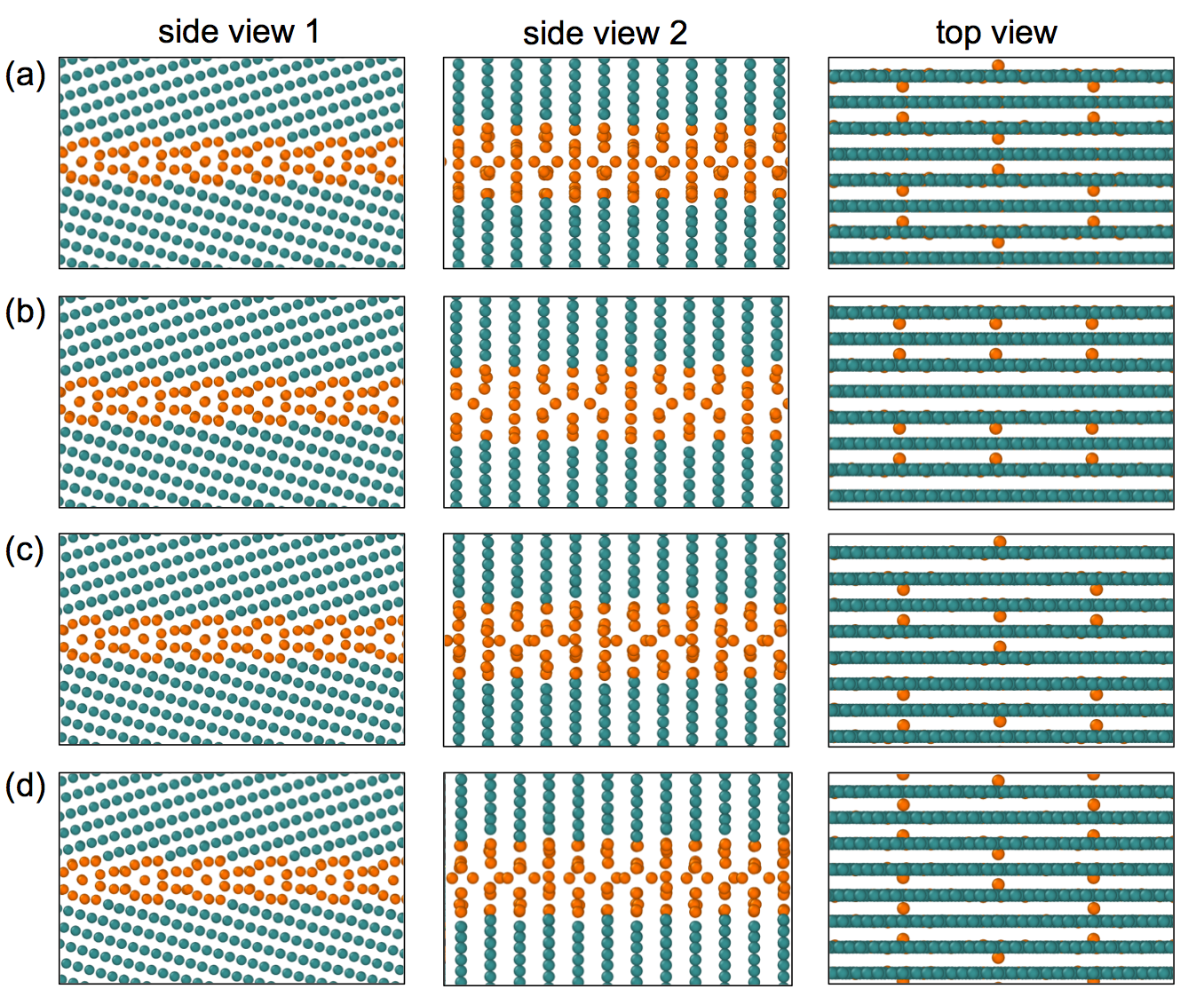} 
\par\end{centering}

\protect\protect\protect\caption{New {[}n{]}=1/2 ground state structures of the $\Sigma27(552)[1\overline{1}0]$
GB in W predicted by the evolutionary structure search with the EAM1
\cite{Marinica2013} and EAM2 \cite{Zhou2001} potentials: (a) GB11,
(b) GB1, (c) GB2 and (d) GB3. The DFT calculations confirm these to
be the lowest energy states with the energies $\gamma_{\text{GB}}=2.59$
J/m$^{2}$ identical within the accuracy of the calculations. Bulk
(green) and grain boundary (orange) atoms are colored according to
the common neighbor analysis \cite{0965-0393-18-1-015012}. \label{fig:USPEX}}
\end{figure}

\begin{figure}
\begin{centering}
\includegraphics[width=0.8\paperwidth]{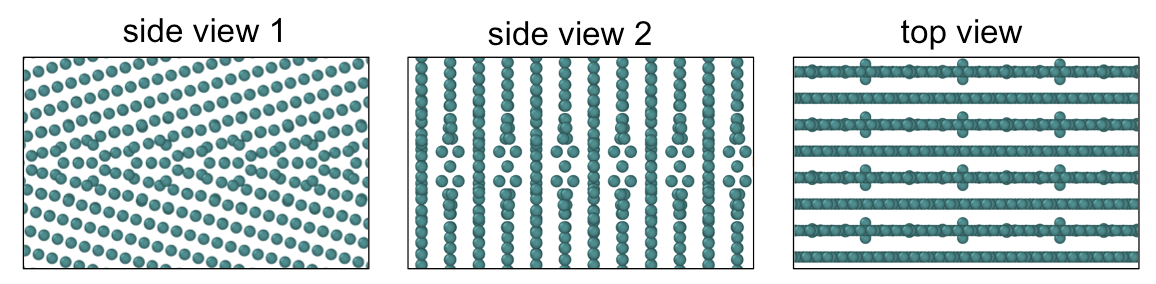} 
\par\end{centering}

\protect\protect\protect\caption{GB12 structure of the $\Sigma27(552)[1\overline{1}0]$ GB in W predicted
by the evolutionary search at $[n]_{\text{GB12}}=0$ with the EAM2
potential \cite{Zhou2001}. The energy of this structure $\gamma_{\text{GB12}}=2.495$
J/m$^{2}$ is identical within the accuracy of the calculations to
the energy of the ground state structure GB11 with $\gamma_{\text{GB11}}=2.493$
J/m$^{2}$ and $[n]_{\text{GB11}}=1/2$ (cf.~Fig.~\ref{fig:USPEX}a).
This structure has the same number of atoms as the $\gamma$-surface
structure, but the energy is 7\% lower. In this case no atoms were
inserted or removed from the GB core, however the evolutionary search
finds this low-energy structure by rearranging the GB atoms. \label{fig:Zhou_gb_phase_n0}}
\end{figure}

Figures \ref{fig:USPEX-search_GB_energy} a and b illustrate the results
of the evolutionary searches performed using the EAM1 and EAM2 potentials,
respectively. Each blue circle on the plot corresponds to a grain
boundary structure generated during the search. The grain boundary
energy is plotted as a function of the number of atoms {[}n{]} expressed
as a fraction of atoms in the bulk (552) plane. The red diamonds on
the plots represent the best configurations generated by the $\gamma$-surface
approach. At {[}n{]}=1/2 the search with both potentials predicted
new ground states of this boundary with energies 7-12\% (depending
on the potential) lower than that of the $\gamma$-surface generated
structures. Figure \ref{fig:USPEX} illustrates the structures of
several {[}n{]}=1/2 GBs predicted by the two potentials. To obtain
these ground states, a number of atoms equal to half (1/2) of the
(552) atomic plane must be inserted into the GB core. This explains
why these structures have not been discovered by the $\gamma$-surface
method. The low-energy states represent $1\times2$ and $1\times3$
unit cell reconstructions and cannot be found in a standard $1\times1$
unit cell. At {[}n{]}=0 the EAM2 potential~\cite{Zhou2001} also
predicts a different low-energy GB structure, labeled as GB12, that
does not require addition or removal of atoms. The boundary structure
is illustrated in Fig. \ref{fig:Zhou_gb_phase_n0}. The energy of
this structure is nearly the same energy as the new ground state at
the $[n]=1/2$ atomic fraction and represents what may be a different
phase of this boundary. The search with EAM2 potential clearly demonstrates
that insertion or removal of atoms is not the only shortcoming of
the $\gamma$-surface method: even at {[}n{]}=0 there may be distinct
local minima, so prediction of grain boundary structure requires advanced
sampling of many possible configurations.

The new ground states predicted by the evolutionary search are not
unique. The searches with the two potentials identified about thirty
distinct low-energy configurations all within a 2\% energy range (approximately
0.05 J/m$^{2}$). The energies of the best thirteen configurations
were subsequently refined with DFT calculations. See the Supplemental
Material \cite{suplTF1/online} for the details of the DFT calculations.
The results of the calculations are summarized in Table~1. The DFT
calculations confirmed that {[}\textit{n}{]}=1/2 ground states predicted
by both potentials have essentially the same energy. Figure~\ref{fig:USPEX}
illustrates four examples of the ground state structures. These boundaries
correspond to GBs 11, 1, 2 and 3 in Table 1. The structure in Fig.~\ref{fig:USPEX}a
was predicted by potential EAM2, while the other structures were generated
using the the EAM1 potential. Each GB structure is shown from three
different viewpoints. The left-hand side panel shows that all the
structures are nearly indistinguishable when viewed projected on the
plane normal to the $[1\overline{1}0]$ tilt axis, the standard view
to visualize structural units. On the other hand, the middle and the
right-hand panels show significantly different atomic arrangements.
In the middle panels the tilt axis is parallel to the plane of the
image, while the right-hand side panels show the structure within
the GB plane viewed from the top. The top view in the right-hand side
panels most clearly shows the difference between these ordered structures.
In all of these configurations the atoms occupy sites between the
$(1\overline{1}0)$ planes within the GB plane. These calculations
show that a number of similar structures with the same GB energy can
be generated by permuting the occupancy of atoms in different \textit{interstitial}
positions within the boundary. These structures can no longer be mapped
onto a DSC lattice, and to distinguish them from the structures generated
by the $\gamma$-surface approach we refer to them as non-DSC structures.
This characteristic feature is remarkably similar to split-kite phases
recently found in {[}100{]} symmetric tilt boundaries in Cu~\cite{Frolov2013},
suggesting that these non-DSC structures may be a general phenomenon.

At {[}n{]}=0 the predictions of the three models are less consistent.
The DFT energies of the $\gamma$-surface structures GB10 (2.960~J/m$^{2}$)
and GB13 (2.973~J/m$^{2}$) generated with the empirical potentials
do not agree with the energy of the $\gamma$-surface structure GB14
(2.688~J/m$^{2}$) based only on DFT. The GB12 ground state at {[}n{]}=0
predicted by the evolutionary search with EAM2 potential was also
not confirmed as a low energy state by the subsequent DFT calculations.
It is possible that the evolutionary search that uses DFT calculations
only would generate yet a different low-energy state at {[}n{]}=0.
Unfortunately, such a calculation would be significantly more expensive.
Nevertheless, within the DFT model the structures GB14 at {[}n{]}=0
with energy $\gamma_{\text{GB14}}=2.68$ J/m$^{2}$ and GB1 at {[}n{]}=1/2
with energy $\gamma_{\text{GB1}}=2.592$ J/m$^{2}$ represent two
candidates for distinct grain boundary phases. The close energies
at 0 K suggest the possibility of transitions between the two GB structures
due to temperature, pressure or alloying.

\begin{figure}
\begin{centering}
\includegraphics[height=0.6\paperheight]{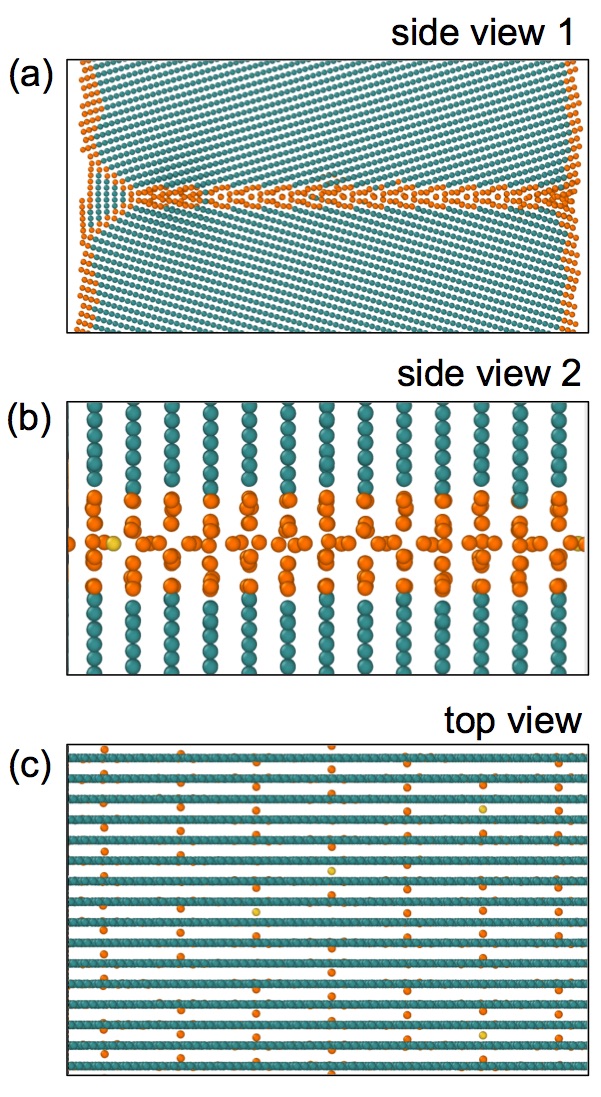} 
\par\end{centering}

\protect\protect\protect\caption{Equilibrium structure of the $\Sigma27(552)[1\overline{1}0]$ GB at
T=2500 K modeled with the EAM1 potential. A 100 ns long isothermal
molecular dynamic simulation with the GB terminated at two open surfaces
predicts the high-temperature GB stucture independently from the 0
K search. Bulk (green) and grain boundary (orange) atoms are colored
according to the common neighbor analysis \cite{0965-0393-18-1-015012}.
\label{fig:High T}}
\end{figure}

The large number of GB structures nearly degenerate in energy found
by the evolutionary search at 0 K suggests new questions about tungsten
grain boundaries at finite temperature. Can some of the different
grain boundary structures coexist in equilibrium? How does the multiplicity
of similar {[}n{]}=1/2 structures affect the finite-temperature structure?
The abundance of similar structures may contribute to the configurational
entropy of the boundary at finite temperature~\cite{Han:2016aa},
since many different states can be created by permutations of atoms
in different sites within the boundary with a negligible penalty in
energy. To investigate the effect of temperature on {[}n{]}=1/2 GB
structure, we performed a molecular dynamics simulation for 100 ns
at high temperature (2500~K) with the EAM1 potential. To avoid bias
for one of the newly identified {[}\textit{n}{]}=1/2 ground states,
we used a higher-energy structure predicted by the $\gamma$-surface
approach (Fig.~\ref{fig:Kites_gamma_surface_approach}b) as the initial
configuration. Along the $x$ direction we terminated the GB with
two $(\overline{1}15)$ surfaces to allow atoms to diffuse in and
out, enabling the atomic density in the GB core to vary~\cite{Frolov2013}.
During the simulation, the GB transforms to its non-DSC state. The
equilibrium high-temperature structure is illustrated in Fig.~\ref{fig:High T}.
The simulation confirms that the non-DSC ground state identified by
the evolutionary search remains the minimum free energy structure
at high temperature. The examination of the boundary structure viewed
from the top in Fig.~\ref{fig:High T}b reveals that the high-temperature
structure is a combination of different structures shown in Figs.~\ref{fig:USPEX}c
and d \cite{Kurtz02,Han:2016aa}. The left surface shows a $chevron$
reconstruction~\cite{PhysRevB.69.172102,PhysRevLett.89.085502}.
Near the chevron the first two GB units have a different structure
closely resembling the {[}n{]}=0 non-DSC ground state GB12 identified
using the EAM2 potential. This example also shows that the GB structure
in a polycrystalline metal will be influenced by local mechanical
forces (e.g. triple junctions, GB defects, nearby lattice dislocations,
etc.)

\begin{figure}
\begin{centering}
\includegraphics[width=0.8\paperwidth]{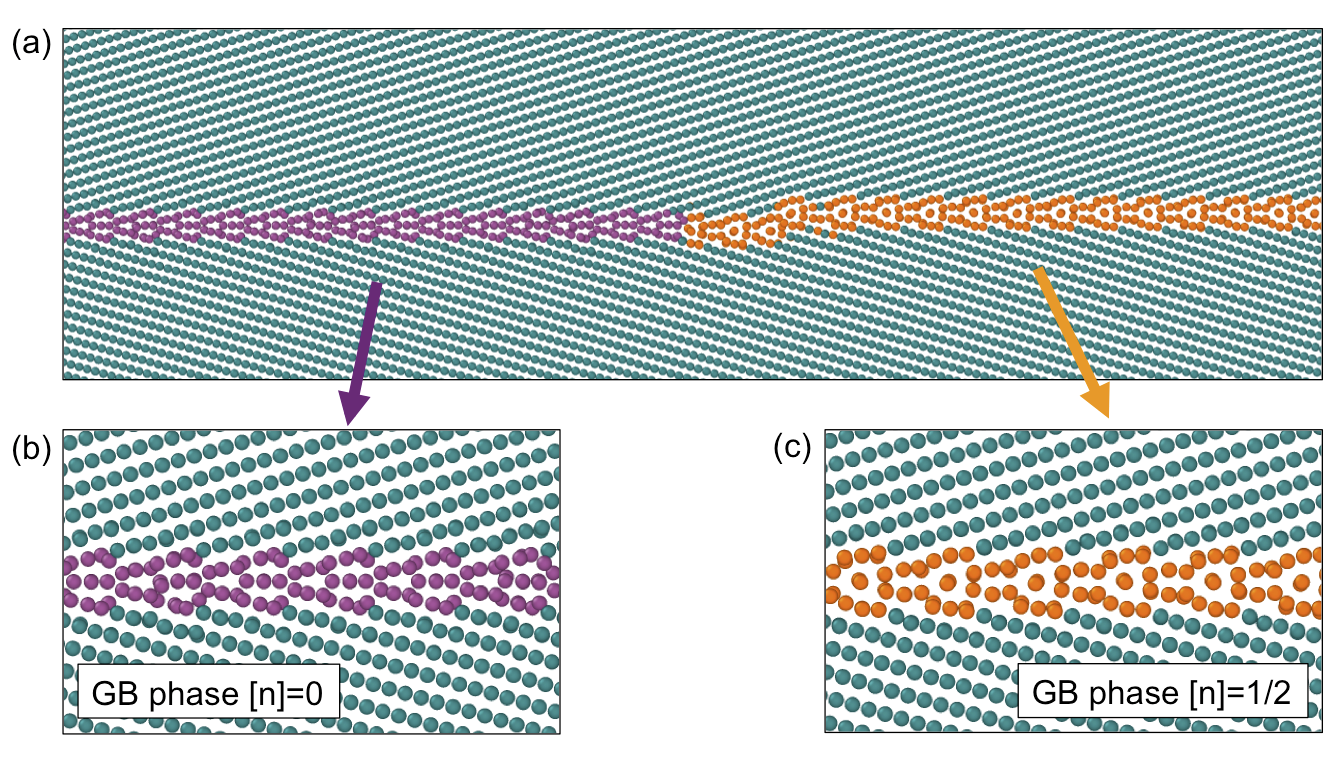} 
\par\end{centering}

\protect\protect\protect\caption{(a) Two grain boundary phases {[}n{]}=0 and {[}n{]}=1/2 coexisting
in equilibrium at T=2000 K for 200 ns in a closed system with periodic
boundary conditions along the boundary. b) and c) show closer views
of the two structures. The equilibrium is established through the
exchange of atoms bethween the two GB phases. The coexistence simulation
demonstrates that the two types of structres predicted by the evolutionary
search at 0 K represent two different GB phases. The $\gamma$-surface
approach fails to predict both of these finite-temperature structures.
In the figure the bulk (green) and the grain boundary (orange, magenta)
atoms are colored according to the common neighbor analysis \cite{0965-0393-18-1-015012}.
\label{fig:Phase-coexistence}}
\end{figure}

The evolutionary search with the EAM2 potential predicts two distinct
low-energy structures with {[}n{]}=0 and {[}n{]}=1/2. MD simulations
of the individual structures at T=2000 K and T=2500 K with periodic
boundary conditions confirmed that both are stable at finite temperature.
To test whether the two types of structures can coexist we created
a simulation block with dimensions $49.5\times2.7\times13.0$ nm$^{3}$
and periodic boundary conditions along the boundary. The initial GB
structure was set to GB12. Then, additional atoms were inserted at
random positions in one half of the bicrystal at a distance 5 to 10~{\AA }
above the GB plane. The number of atoms inserted in that section of
the block was equal to half of a (552) atomic plane. This configuration
was annealed at 2000~K for 200 ns. During the first few nanoseconds
of the simulation the added atoms diffused into the GB and about half
of the total GB area transformed into the {[}n{]}=1/2 structure. After
that, the two grain boundary structures continued to coexist for the
rest of the simulation time and we observed no further transformations.
Figure \ref{fig:Phase-coexistence} illustrates the {[}n{]}=0 and
{[}n{]}=1/2 grain boundary phases coexisting in equilibrium. The two
structures are separated by a line defect that spans the periodic
length of the simulation block. The position of this line defect fluctuates
during the simulation. The equilibrium in this closed system with
periodic boundary conditions is established by exchange of atoms diffusing
along the boundary. The coexistence simulation demonstrate that the
two types of structures predicted by the evolutionary search represent
two GB phases. The transformation is first-order and results in a
discontinuous change in excess GB properties. This is to be contrasted
to higher order transitions such as continuous premelting when only
one GB state can exist at given temperature and pressure. To the best
of our knowledge this is a first demonstration of phase behavior of
high-angle GBs in a bcc material.

In conclusion, conventional simulation methodologies such as the $\gamma$-surface
method often predict relatively simple structures for symmetric tilt
boundaries, which are composed of kite-shaped units such as illustrated
in Fig.~\ref{fig:Kites_gamma_surface_approach}. Their structure
can be described within the structural unit model which is based on
bulk crystallography~\cite{Han2017}. For this reason symmetric tilt
boundaries are considered to be some of the simplest boundaries, and
they are popular model systems. In this work we have demonstrated
that in bcc material such as tungsten, the structure of symmetric
tilt boundaries can be significantly more complex.

We performed a grand-canonical evolutionary structure search of the
$\Sigma27(552)[1\overline{1}0]$ boundary; i.e. a search in which
the number of atoms can vary. This boundary has been investigated
in the past to study solute segregation and GB embrittlement~\cite{Kurtz2014,SetJPCM14}.
Our calculations with two different interatomic potentials and DFT
predict a new ground state, which requires additional atoms equivalent
to half of a $(552)$ atomic plane. The lack of atomic density optimization
and the absence of sufficient sampling are the two main reasons this
ground state was not found previously.

The new ground state structures are characterized by complex arrangement
of atoms within the GB plane. The boundaries are composed of a number
of atoms incompatible with the number of atoms per atomic plane in
the abutting grains. The GB structure cannot be mapped onto the DSC
lattice. The ground state is degenerate, represented by a large number
of similar structures with the same energy. This configurational complexity
has consequences for the finite-temperature GB structure, which we
observe to be comprised of a combination of states found at 0 K. The
structural features are remarkably similar to split-kite phases found
in symmetric tilt fcc GBs~\cite{Frolov2013,qzhutfrolov/online}.

Within the EAM2 model the evolutionary search at 0 K identified two
distinct low-energy GB structures with different atomic densities
{[}n{]}=0 and {[}n{]}=1/2. High-temperature MD simulations demonstrated
that the two structures can coexist in equilibrium in a closed system
while exchanging atoms by diffusion. This simulation confirms that
the two structures are examples of two GB phases. Within the DFT model
the energy difference between the different structures GB14 and GB1
is only 3\%. The closeness of the energies at 0 K also suggests a
possibility of transitions between the two GB structures due to temperature,
pressure or addition of solute atoms.

Transformations at grain boundaries are not only of fundamental scientific
interest, but may also have practical importance by affecting the
properties of materials. A number of recent experimental studies demonstrated
discontinuous changes in properties of polycrystalline materials and
bicrystals, linking grain boundary phase transitions to abnormal grain
growth, activated sintering and grain boundary embrittlement~\cite{Baram08042011,Luo23092011,Harmer08042011,Rheinheimer201568,Cantwell20141,SCHULER2017196,Rohrer2016231}.
Multiple GB phases found by atomistic simulations in fcc Cu provided
a convenient model to investigate the importance of GB structure-property
relations. Specifically, the simulations revealed that the transitions
between these GB structures have a pronounced effect on shear strength
and can even reverse the direction of GB migration~\cite{doi:10.1063/1.4880715,Borovikov2013}.
In a binary Cu(Ag) system, the different GB phases demonstrated distinct
monolayer and bilayer segregation patterns with very different amounts
of Ag segregation~\cite{PhysRevB.92.020103,Frolov2013PRL}. In other
words, the changes in GB structure can dramatically alter the segregation
sites and the occupation of these sites by solutes.

In this work we demonstrated new ground states and phase behavior
of grain boundaries in a model bcc metal. In our high-temperature
simulations the GB transition (the nucleation of the second GB phase)
was triggered by absorption of interstitial atoms. This mechanism
of defect accommodation by GBs may be relevant to higher radiation
tolerance of nanocrystalline materials. The detailed investigation
of impurity segregation to non-DSC grain boundaries and their mechanical
properties is the subject of future work. The rich behavior found
in a high-angle and high-energy $\Sigma27(552)[1\overline{1}0]$ grain
boundary using the new evolutionary method motivates a systematic
investigation of other grain boundaries in bcc metals as well as grain
boundary phase transitions in tungsten alloys.

\begin{table}
\begin{centering}
\begin{tabular}{|c|c|c|c|c|}
\hline 
Label  & {[}\textit{n}{]}  & EAM1, J/m$^{2}$  & EAM2, J/m$^{2}$  & DFT, J/m$^{2}$\tabularnewline
\hline 
\hline 
GB1  & 1/2  & 2.819  &  & 2.592\tabularnewline
\hline 
GB2  & 1/2  & 2.811  &  & 2.593\tabularnewline
\hline 
GB3  & 1/2  & 2.818  &  & 2.594\tabularnewline
\hline 
GB4  & 1/2  & 2.807  &  & 2.595\tabularnewline
\hline 
GB5  & 1/2  & 2.817  &  & 2.609\tabularnewline
\hline 
GB6  & 1/2  & 2.802  &  & 2.610\tabularnewline
\hline 
GB7  & 1/2  & 2.798  &  & 2.624\tabularnewline
\hline 
GB8  & 1/2  & 2.796  &  & 2.626\tabularnewline
\hline 
GB9  & 1/2  & 2.812  &  & 2.628\tabularnewline
\hline 
GB10{*}  & 0  & 3.171  &  & 2.960\tabularnewline
\hline 
GB11  & 1/2  &  & 2.493  & 2.590\tabularnewline
\hline 
GB12  & 0  &  & 2.495  & 2.951\tabularnewline
\hline 
GB13{*}  & 0  &  & 2.670  & 2.973\tabularnewline
\hline 
GB14{*}  & 0  &  &  & 2.680\tabularnewline
\hline 
\end{tabular}
\par\end{centering}

\protect\protect\protect\caption{The grain boundary energy of different structures generated with the
evolutionary algorithm and the $\gamma$-surface approach ({*}) using
the EAM1 and EAM2 potentials and DFT. The second column indicates
the atomic density $[n]$ of the different structures as a fraction
of the atoms in the (552) plane. Both potentials predict new ground
states that were found with the grand-canonical evolutionary search.
The ground state is represented by several similar but distinct structures
with the same energy within the accuracy of the DFT calculations.
GB14 was previously found in Ref. \cite{SetJPCM14}.}
\end{table}

\section*{Acknowledgment}

This work was performed under the auspices of the U.S. Department
of Energy (DOE) by Lawrence Livermore National Laboratory under contract
DE-AC52-07NA27344 and by Pacific Northwest National Laboratory under
contract DE-AC05-76RLO-1830. This material is based upon work supported
by the U.S. DOE, Office of Science, Office of Fusion Energy Sciences.
The work was supported by the Laboratory Directed Research and Development
Program at LLNL, project 17-LW-012. We acknowledge the use of LC computing
resources. Work in UNLV is supported by the National Nuclear Security
Administration under the Stewardship Science Academic Alliances program
through DOE Cooperative Agreement DE-NA0001982.

\begin{thebibliography}{10}

\bibitem{Balluffi95}
A.~P. Sutton and R.~W. Balluffi,
\newblock {\em Interfaces in Crystalline Materials},
\newblock Clarendon Press, Oxford, 1995.

\bibitem{JACE:JACE03545}
M.~P. Harmer,
\newblock Journal of the American Ceramic Society {\bf 93}, 301 (2010).

\bibitem{Luo23092011}
J.~Luo, H.~Cheng, K.~M. Asl, C.~J. Kiely, and M.~P. Harmer,
\newblock Science {\bf 333}, 1730 (2011).

\bibitem{tagkey2004xxi}
Symbols,
\newblock in {\em Recrystallization and Related Annealing Phenomena (Second
  Edition)}, edited by F.~Humphreys and M.~Hatherly, pages xxi -- xxii,
  Elsevier, Oxford, second edition edition, 2004.

\bibitem{Lu349}
K.~Lu, L.~Lu, and S.~Suresh,
\newblock Science {\bf 324}, 349 (2009).

\bibitem{doi:10.1146/annurev-matsci-070813-113627}
S.~Zinkle and L.~Snead,
\newblock Annual Review of Materials Research {\bf 44}, 241 (2014).

\bibitem{0029-5515-47-11-014}
D.~Maisonnier, D.~Campbell, I.~Cook, L.~D. Pace, L.~Giancarli, J.~Hayward,
  A.~L. Puma, M.~Medrano, P.~Norajitra, M.~Roccella, P.~Sardain, M.~Tran, and
  D.~Ward,
\newblock Nuclear Fusion {\bf 47}, 1524 (2007).

\bibitem{Mutoh95}
Y.~Mutoh, K.~Ichikawa, K.~Nagata, and M.~Takeuchi,
\newblock J. of Mat. Sci. {\bf 30}, 770 (1995).

\bibitem{MillerMSEA02}
M.~K. Miller and A.~J. Bryhan,
\newblock Materials Science and Engineering: A {\bf 327}, 80 (2002).

\bibitem{NiehSM84}
T.~G. Nieh,
\newblock Scripta Metallurgica {\bf 18}, 1279 (1984).

\bibitem{RICE198923}
J.~R. Rice and J.-S. Wang,
\newblock Materials Science and Engineering: A {\bf 107}, 23  (1989).

\bibitem{SetSM12}
W.~Setyawan and R.~J. Kurtz,
\newblock Scripta Materialia {\bf 66}, 558 (2012).

\bibitem{SetJPCM14}
W.~Setyawan and R.~J. Kurtz,
\newblock Journal of Physics: Condensed Matter {\bf 26}, 135004 (2014).

\bibitem{ScheiAM15}
D.~Scheiber, V.~I. Razumovskiy, P.~Puschnig, R.~Pippan, and L.~Romaner,
\newblock Acta Materialia {\bf 88}, 180 (2015).

\bibitem{WuAM16}
X.~Wu, Y.-W. You, X.-S. Kong, J.-L. Chen, G.~N. Luo, G.-H. Lu, C.~S. Liu, and
  Z.~Wang,
\newblock Acta Materialia {\bf 120}, 315 (2016).

\bibitem{PSSA:PSSA2211180112}
G.~Schoeck and W.~Pichl,
\newblock Phys. Status Solidi A {\bf 118}, 109 (1990).

\bibitem{Moller20141}
J.~J. Moller and E.~Bitzek,
\newblock Acta Materialia {\bf 73}, 1  (2014).

\bibitem{MORITA19971053}
K.~Morita and H.~Nakashima,
\newblock Materials Science and Engineering: A {\bf 234}, 1053  (1997).

\bibitem{PhysRevB.85.064108}
M.~A. Tschopp, K.~N. Solanki, F.~Gao, X.~Sun, M.~A. Khaleel, and M.~F.
  Horstemeyer,
\newblock Phys. Rev. B {\bf 85}, 064108 (2012).

\bibitem{doi:10.1080/13642818908211183}
D.~Wolf,
\newblock Philosophical Magazine Part B {\bf 59}, 667 (1989).

\bibitem{doi:10.1063/1.347741}
D.~Wolf,
\newblock Journal of Applied Physics {\bf 69}, 185 (1991).

\bibitem{PhysRevB.64.174101}
D.~Ye\ifmmode~\mbox{\c{s}}\else \c{s}\fi{}illeten and T.~A. Arias,
\newblock Phys. Rev. B {\bf 64}, 174101 (2001).

\bibitem{Ratanaphan2015346}
S.~Ratanaphan, D.~L. Olmsted, V.~V. Bulatov, E.~A. Holm, A.~D. Rollett, and
  G.~S. Rohrer,
\newblock Acta Materialia {\bf 88}, 346  (2015).

\bibitem{doi:10.1080/01418618308243118}
P.~W. Tasker and D.~M. Duffy,
\newblock Philos. Mag. A {\bf 47}, L45 (1983).

\bibitem{doi:10.1080/01418618608242811}
D.~M. Duffy and P.~W. Tasker,
\newblock Philos. Mag. A {\bf 53}, 113 (1986).

\bibitem{DUFFY84a}
D.~M. Duffy and P.~W. Tasker,
\newblock J. Am. Ceram. Soc {\bf 67}, 176 (1984).

\bibitem{Phillpot1992}
S.~R. Phillpot and J.~M. Rickman,
\newblock The Journal of Chemical Physics {\bf 97}, 2651 (1992).

\bibitem{Alfthan06}
S.~\mbox{von Alfthan}, P.~D. Haynes, K.~Kashi, and A.~P. Sutton,
\newblock Phys. Rev. Lett. {\bf 96}, 055505 (2006).

\bibitem{PhysRevB.80.174102}
J.~Zhang, C.-Z. Wang, and K.-M. Ho,
\newblock Phys. Rev. B {\bf 80}, 174102 (2009).

\bibitem{Chua:2010uq}
A.~L.~S. Chua, N.~A. Benedek, L.~Chen, M.~W. Finnis, and A.~P. Sutton,
\newblock Nat Mater {\bf 9}, 418 (2010).

\bibitem{Frolov2013}
T.~Frolov, D.~L. Olmsted, M.~Asta, and Y.~Mishin,
\newblock Nat. Commun. {\bf 4}, 1899 (2013).

\bibitem{Frolov2013PRL}
T.~Frolov, S.~V. Divinski, M.~Asta, and Y.~Mishin,
\newblock Phys. Rev. Lett. {\bf 110}, 255502 (2013).

\bibitem{qzhutfrolov/online}
{Q. Zhu, A. Samanta, B. Li, R. E. Rudd, and T. Frolov, arxiv:1707.09699.}

\bibitem{Frolov2016}
T.~Frolov, M.~Asta, and Y.~Mishin,
\newblock Curr. Opin. Solid State Mater. Sci. {\bf 20}, 308  (2016).

\bibitem{PhysRevB.92.020103}
T.~Frolov, M.~Asta, and Y.~Mishin,
\newblock Phys. Rev. B {\bf 92}, 020103 (2015).

\bibitem{Demkowicz2015}
W.~Yu and M.~Demkowicz,
\newblock Journal of Materials Science {\bf 50}, 4047 (2015).

\bibitem{Novoselov2016276}
I.~Novoselov and A.~Yanilkin,
\newblock Computational Materials Science {\bf 112, Part A}, 276  (2016).

\bibitem{Han2017}
J.~Han, V.~Vitek, and D.~J. Srolovitz,
\newblock Acta Materialia {\bf 133}, 186  (2017).

\bibitem{Han:2016aa}
J.~Han, V.~Vitek, and D.~J. Srolovitz,
\newblock Acta Mater. {\bf 104}, 259 (2016).

\bibitem{Olmsted2011}
D.~L. Olmsted, D.~Buta, A.~Adland, S.~M. Foiles, M.~Asta, and A.~Karma,
\newblock Phys. Rev. Lett. {\bf 106}, 046101 (2011).

\bibitem{Reilly:gp5080}
A.~M. Reilly et~al.,
\newblock Acta Crystallographica Section B {\bf 72}, 439 (2016).

\bibitem{doi:10.1063/1.2210932}
A.~R. Oganov and C.~W. Glass,
\newblock The Journal of Chemical Physics {\bf 124}, 244704 (2006).

\bibitem{Zhou-PRL-2014}
X.-F. Zhou, X.~Dong, A.~R. Oganov, Q.~Zhu, Y.~Tian, and H.-T. Wang,
\newblock Phys. Rev. Lett. {\bf 112}, 085502 (2014).

\bibitem{Zhu-PRB-2013}
Q.~Zhu, L.~Li, A.~R. Oganov, and P.~B. Allen,
\newblock Phys. Rev. B {\bf 87}, 195317 (2013).

\bibitem{Zhu-JCP-2014}
Q.~Zhu, V.~Sharma, A.~R. Oganov, and R.~Ramprasad,
\newblock The Journal of Chemical Physics {\bf 141}, 154102 (2014).

\bibitem{Lyakhov-CPC-2013}
A.~O. Lyakhov, A.~R. Oganov, H.~T. Stokes, and Q.~Zhu,
\newblock Computer Physics Communications {\bf 184}, 1172  (2013).

\bibitem{doi:10.1021/ar1001318}
A.~R. Oganov, A.~O. Lyakhov, and M.~Valle,
\newblock Accounts of Chemical Research {\bf 44}, 227 (2011).

\bibitem{ScheiMSMSE16}
D.~Scheiber, R.~Pippan, P.~Puschnig, and L.~Romaner,
\newblock Modelling and Simulation in Materials Science and Engineering {\bf
  24}, 085009 (2016).

\bibitem{Li14}
Z.~W. Li, X.~S. Kong, Liu-Wei, C.~S. Liu, and Q.~F. Fang,
\newblock Chinese Physics B {\bf 23}, 106107 (2014).

\bibitem{Marinica2013}
M.~C. Marinica, L.~Ventelon, M.~R. Gilbert, L.~Proville, S.~L. Dudarev,
  J.~Marian, G.~Bencteux, and F.~Willaime,
\newblock Journal of Physics: Condensed Matter {\bf 25}, 395502 (2013).

\bibitem{Zhou2001}
X.~Zhou, H.~Wadley, R.~Johnson, D.~Larson, N.~Tabat, A.~Cerezo,
  A.~Petford-Long, G.~Smith, P.~Clifton, R.~Martens, and T.~Kelly,
\newblock Acta Materialia {\bf 49}, 4005  (2001).

\bibitem{Kurtz2014}
W.~Setyawan and R.~J. Kurtz,
\newblock Journal of Physics: Condensed Matter {\bf 26}, 135004 (2014).

\bibitem{Zhu-PRB-2015}
Q.~Zhu, A.~R. Oganov, A.~O. Lyakhov, and X.~Yu,
\newblock Phys. Rev. B {\bf 92}, 024106 (2015).

\bibitem{Plimpton95}
S.~Plimpton,
\newblock J. Comput. Phys. {\bf 117}, 1 (1995).

\bibitem{LYAKHOV20101623}
A.~O. Lyakhov, A.~R. Oganov, and M.~Valle,
\newblock Comput. Phys. Commun. {\bf 181}, 1623  (2010).

\bibitem{suplTF1/online}
See supplemental material at [url] for complete results of the evolutionary
  search search, the details of the dft calculations and the atomic structure
  of {GB}12.

\bibitem{Kurtz02}
R.~G. Hoagland and R.~J. Kurtz,
\newblock Philos. Mag. {\rm A} {\bf 82}, 1073 (2002).

\bibitem{PhysRevB.69.172102}
F.~Lan\ifmmode~\mbox{\c{c}}\else \c{c}\fi{}on, T.~Radetic, and U.~Dahmen,
\newblock Phys. Rev. B {\bf 69}, 172102 (2004).

\bibitem{PhysRevLett.89.085502}
T.~Radetic, F.~Lan\ifmmode~\mbox{\c{c}}\else \c{c}\fi{}on, and U.~Dahmen,
\newblock Phys. Rev. Lett. {\bf 89}, 085502 (2002).

\bibitem{Baram08042011}
M.~Baram, D.~Chatain, and W.~D. Kaplan,
\newblock Science {\bf 332}, 206 (2011).

\bibitem{Harmer08042011}
M.~P. Harmer,
\newblock Science {\bf 332}, 182 (2011).

\bibitem{Rheinheimer201568}
W.~Rheinheimer and M.~J. Hoffmann,
\newblock Scripta Materialia {\bf 101}, 68  (2015).

\bibitem{Cantwell20141}
P.~R. Cantwell, M.~Tang, S.~J. Dillon, J.~Luo, G.~S. Rohrer, and M.~P. Harmer,
\newblock Acta Materialia {\bf 62}, 1  (2014).

\bibitem{SCHULER2017196}
J.~D. Schuler and T.~J. Rupert,
\newblock Acta Materialia {\bf 140}, 196  (2017).

\bibitem{Rohrer2016231}
G.~S. Rohrer,
\newblock Curr. Opin. Solid State Mater. Sci. {\bf 20}, 231  (2016).

\bibitem{doi:10.1063/1.4880715}
T.~Frolov,
\newblock Appl. Phys. Lett. {\bf 104}, 211905 (2014).

\bibitem{Borovikov2013}
V.~Borovikov, X.-Z. Tang, D.~Perez, X.-M. Bai, B.~P. Uberuaga, and A.~F. Voter,
\newblock Journal of Physics: Condensed Matter {\bf 25}, 035402 (2013).

\bibitem{0965-0393-18-1-015012}
A.~Stukowski,
\newblock Modell. Simul. Mater. Sci. Eng. {\bf 18}, 015012 (2010).

\end{thebibliography}
\end{document}